\documentclass[aps,prl,twocolumn]{revtex4}
\usepackage{amssymb}
\usepackage{amsfonts}
\usepackage{amsmath}
\usepackage{amsmath}
\usepackage{amssymb}
\usepackage{graphicx}

\begin{document}

\title{Spin-orbit coupling and anisotropy of spin splitting in quantum dots}
\author{J. K\"{o}neman$^{1}$, R.J. Haug$^{1}$, D. K. Maude$^{2}$, V.I. Fal'ko$%
^{3,4}$, B.L. Altshuler$^{4}$}
\affiliation{$^{1}$ Institut f\"{u}r Festk\"{o}rperphysik, Universit\"{a}t Hannover,
Appelstrasse 2, D-30167 Hannover, Germany}
\affiliation{$^{2}$ High Magnetic Field Laboratory, CNRS, 25 Avenue des Martyrs, BP 166,
F-38042 Grenoble cedex 9, France}
\affiliation{$^{3}$ Physics Department, Lancaster University, Lancaster LA1 4YB, United
Kingdom}
\affiliation{$^{4}$ Physics Department, Jadwin Hall, Princeton University, NJ 08450}
\date{\today}

\begin{abstract}
In lateral quantum dots, the combined effect of both Dresselhaus and Bychkov-Rashba spin
orbit coupling is equivalent to an effective magnetic field $\pm B_{so}$ which has the
opposite sign for $s_{z}=\pm 1/2$ spin electrons. When the external magnetic field is
perpendicular to the planar structure, the field $B_{so}$ generates an additional
splitting for electron states as compared to the spin splitting in the in-plane field
orientation. The anisotropy of spin splitting has been measured and then analyzed in terms
of spin-orbit coupling in several AlGaAs/GaAs quantum dots by means of resonant tunneling
spectroscopy. From the measured values and sign of the anisotropy we are able to determine
the dominating spin-orbit coupling mechanism.
\end{abstract}

\maketitle

A better understanding of the spin-orbit (SO) effects is crucial for the implementation of the
coherent manipulation of the electron spin \cite%
{SpinManip,SpinQubits} in quantum dots and wires. SO coupling in III-V semiconductor
structures is usually composed of two interplaying
contributions of different symmetries,%
\begin{equation}
\mathcal{H}_{\mathrm{so}}=\varrho _{\mathrm{D}}(p_{x}s_{x}-p_{y}s_{y})+%
\varrho _{\mathrm{BR}}(p_{y}s_{x}-p_{x}s_{y})\mathbf{.}  \label{Hso}
\end{equation}%
The first term is reminiscent of the Dresselhaus  SO coupling in
zink-blend bulk semiconductors \cite{Dress,Dresselhaus} (it
reflects the inversion-asymmetry  of GaAs). The second term in
Eq.~(\ref{Hso}) is the interface-induced coupling of
Bychkov-Rashba type \cite{Rashba}. It is difficult to separate the
effects of the two SO coupling mechanisms  in quantum transport
measurements and spin relaxation
\cite{Rossler,WeakLoc,AF,BF,Zumbuehl,Shayegan}, (except for
optical experiments \cite{Jusserand,Ganichev}), and even to
determine which one is dominant. At the same time coherent spin
manipulation as well as the spin-Hall effect
\cite{SpinManip,SpinManip2,Loss} depend on the balance between the
two mechanisms.

 In this Letter, we show that
the relative strength of the Dresselhaus and Bychkov-Rashba SO
coupling mechanisms in a particular device can be determined from
the anisotropy of the Zeeman spin splitting. Experimentally, we
exploit the method of single-electron resonant tunneling
spectroscopy \cite{Haug} to observe the difference $\Delta _{\bot
}-\Delta _{\Vert }$ in spin splitting of single-electron
resonances in a double-barrier structure subjected to a magnetic
field perpendicular ($\Delta _{\bot }$) and parallel ($\Delta
_{\Vert }$) to the plane of the quantum well. We analyze this
anisotropy within the framework of the theory of SO coupling in
lateral quantum dots \cite{AF,BF}. It is shown below that the two
mechanisms cause anisotropy of opposite signs, and that
\begin{equation}
\Delta _{\bot }-\Delta _{\Vert }\propto \left( -g/|g|\right)
\left( \varrho _{\mathrm{BR}}^{2}-\varrho
_{\mathrm{D}}^{2}\right), \label{diff}
\end{equation}
where $g$ is the quantum well electron Lande g-factor in the
in-plane magnetic field.

 The
experiment was performed with two highly asymmetric double barrier resonant tunneling
devices grown by molecular beam epitaxy on n+-type GaAs substrate. In both samples the
undoped 10~nm wide GaAs quantum well is sandwiched between 5 and 8~nm thick
Al$_{0.3}$Ga$_{0.7}$As tunneling
barriers separated from the highly doped GaAs contacts (Si-doped with $%
n_{S}=4\times 10^{17}$~cm$^{-3}$) by 7~nm thick undoped GaAs spacer layers.
The samples were fabricated as pillars of $2$~$\mu $m (sample A) and $40$~$%
\mu $m diameter (sample B). DC measurements of the I-V characteristics were performed in
two  devices in a dilution refrigerator at $20$~mK base
temperature for two different orientations of magnetic field, see Fig.~\ref%
{fig:sketch} (a).
\begin{figure}
 \center{\includegraphics[width=6cm]{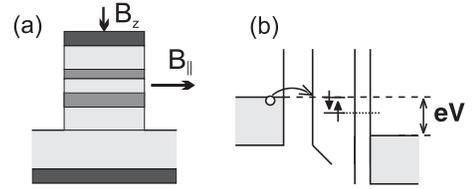}} \caption{Schematic picture
(a) and  energy diagram (b) of the studied device under finite bias and finite magnetic
field.} \label{fig:sketch}
\end{figure}
 The studied GaAs quantum well embedded between two AlGaAs barriers can
be viewed as a two-dimensional system with the edges and residual impurities confining the
lateral electron motion and thus forming dots. Tunneling through the lowest state of the
dot, at the energy $E_{0}$ and with lateral extent $\lambda$, produces the lowest
resonance peak in the differential conductance, whereas its excited states are responsible
for additional peaks in dI/dV, which all move in a magnetic field $B_{z}$ perpendicular to
the quantum well, as shown in Figs. 2 and 3. The magnetic field dependence of energy
levels can be illustrated using the model of parabolic confinement, $V(%
\mathbf{r})=\tfrac{1}{2}m\omega ^{2}\mathbf{r}^{2}$, with the
extension of the wavefunction $\lambda \sim \sqrt{\hbar /\omega
m}$, where the spectrum of quantum dot states $|n_{+}n_{-}\rangle
$, $n_{\pm}=0,1,2,\dots$ is described by
\begin{eqnarray}
E_{n_{+}n_{-}} &=&E_{0}+\hbar \sqrt{\omega ^{2}+\left( \omega _{c}/2\right)
^{2}}-\hbar \omega +\sum_{\pm }n_{\pm }\hbar \omega _{\pm }  \notag \\
\omega _{\pm } &=&\sqrt{\omega ^{2}+\left( \omega _{c}/2\right) ^{2}}\pm
\tfrac{1}{2}\omega _{c},\mbox{ } \omega_c=eB/m.  \label{EnergyLevels}
\end{eqnarray}%
\begin{figure}
\center{\includegraphics[width=7.3cm]{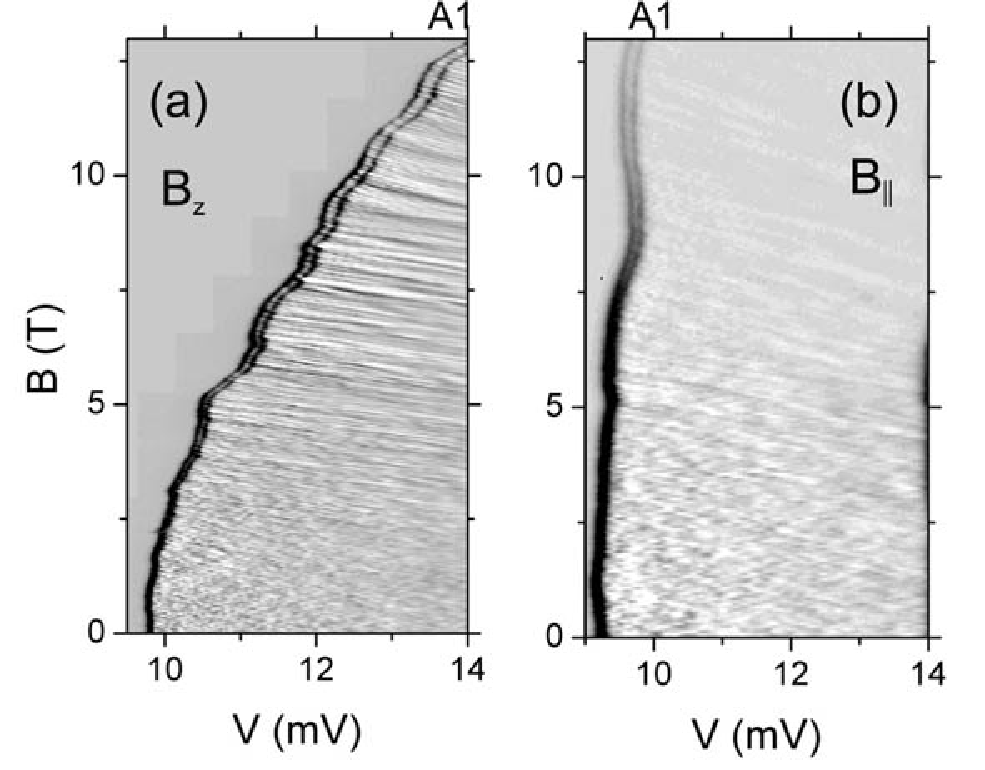}} \caption{Differential
conductance as a grey-scale plot for sample A
($\hbar\omega=31.2$~meV) as a function of bias voltage and
magnetic field for (a) $B_z$ being perpendicular to the quantum
well plane and (b) $B_{\parallel}$, T=20~mK.} \label{fig:idea}
\end{figure}
For a strongly bound state or at low fields, such that $\omega >\omega _{c}$ (also
$\lambda <\lambda _{B}$, where $\lambda _{B}=\sqrt{\hbar /eB_{z}}$), the first tunneling
resonance experiences a diamagnetic shift $E_{00}\approx E_{0}+\hbar \omega
_{c}^{2}/8\omega $, which is quadratic in $B_{z}$. For an electron localized near a smooth
potential minimum, such that $\omega \ll \omega _{c}$ at high fields, $\omega _{+}\approx
\omega _{c}$ and $\omega _{-}\approx \omega ^{2}/\omega _{c}\ll \omega _{c}$, the
diamagnetic shift of several low-energy states in a dot follows approximately the energy
of the lowest 2D Landau level, $E_{0n_{-}}\approx \left[ E_{0}-\hbar \omega
+(n_{-}+1)\hbar \omega ^{2}/\omega _{c}\right] +1/2\hbar \omega _{c}(B_{z})$. \ In the
structures studied, both weakly and strongly bound states have been seen.

\begin{figure}
\center{\includegraphics[width=7.3cm]{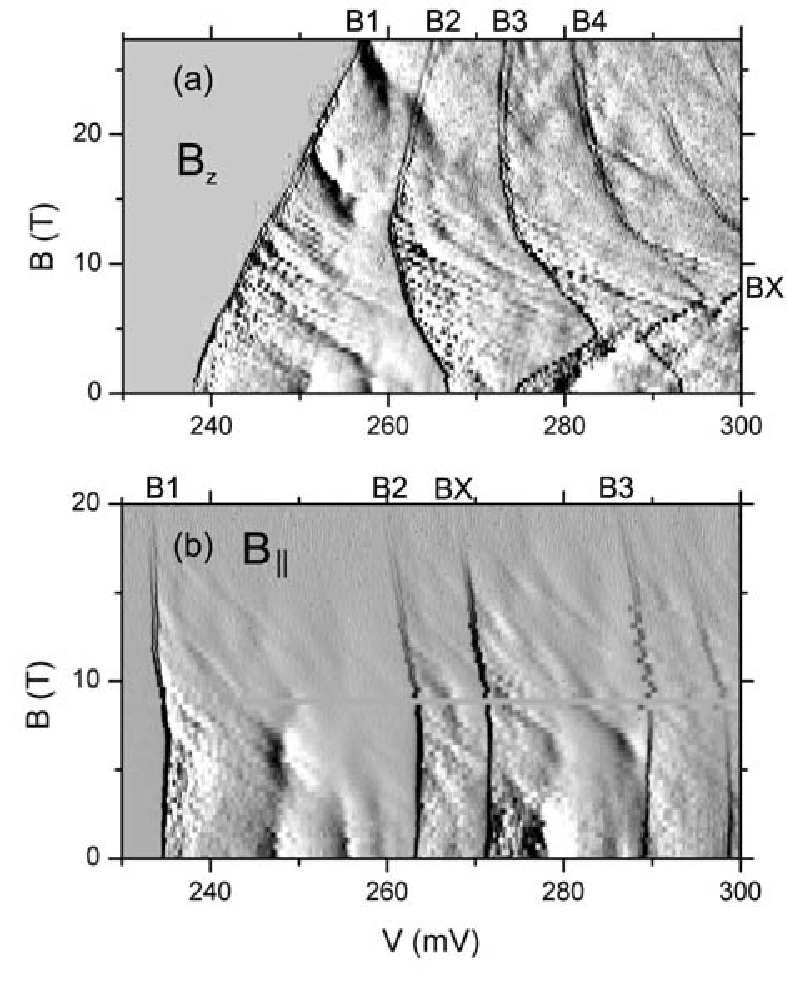}} \caption{Differential
conductance as a grey-scale plot for sample B
($\hbar\omega=13.8$~meV) as a function of bias voltage and
magnetic field for (a) $B_z$ being perpendicular to the quantum
well plane and (b) $B_{\parallel}$, T=20~mK.} \label{fig:idea1}
\end{figure}

Fig.~\ref{fig:idea} (a) shows the differential conductance peak A1 found in sample A and
attributed to a strongly bound state ($\hbar \omega =31.2$~meV), presumably, formed by a
growth-induced local potential minimum. Several differential conductance peaks B1-B4 and
BX were observed in  sample B, with a smaller energy separation. The magnetic field
dependence of their positions shown in Fig.~\ref{fig:idea1}(a) complies with the
magneto-spectrum in a parabolic potential with $\hbar \omega =13.8$~meV \cite{footnote2}.
The data shown in Fig.~\ref{fig:idea} (b) and \ref{fig:idea1}(b), which have been taken on
the same structures subjected to an in-plane magnetic field of comparable strength, do not
display a diamagnetic shift and confirm the 2D nature of the dots.

\begin{figure}
 \center{\includegraphics[width=5.cm]{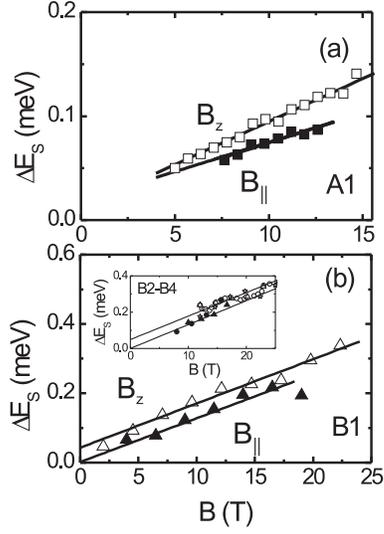}} \caption{(a) Spin
splitting of the state A1 of sample A for both B-field configurations. (b) Spin splitting
of the state B1 of sample B for both B-field configurations. Solid and dotted lines are
 discussed in the text. Inset shows the spin splitting
of the states B2-B4 of sample B for both B-field configurations.} \label{fig:col}
\end{figure}

In both magnetic field directions, all peaks in dI/dV resolve into two at high enough
field values, manifesting the spin splitting of each dot state. Spin splittings extracted
from the set of data shown in Figs.~\ref{fig:idea} and \ref{fig:idea1} are gathered in
Fig.~\ref{fig:col}: (a) shows the data for
 sample A, whereas (b) shows the field dependence of splittings  in  sample B.
   The data
shown in Fig.~\ref{fig:col} display a distinct anisotropy of the peak splitting, with the
splitting caused by the out-of-plane field (open symbols) being systematically larger than
what is created by the in-plane field (closed symbols).

The observed values of spin splitting anisotropy are much larger and have the opposite
sign to what might be expected from the kinetic energy dependence, $g=-0.44+E\cdot dg/dE$
of the electron g-factor across the conduction band in GaAs. Using $dg/dE\sim
+2$~$\mathrm{eV}^{-1}$\cite{Snelling}, we estimate that the diamagnetic shift (which
increases the  electron kinetic energy in a perpendicular magnetic field) would further
reduce the value of $g$ at higher fields by $(\hbar \omega _{c}/2)%
\cdot (dg/dE)$. The latter is less than the observed anisotropy in
the peak B1. One may also notice that the anisotropy of spin
splitting of the excited dot states B2-B4 shown in the inset to
Fig.~\ref{fig:col} is the same as of B1, despite a larger kinetic
energy of an electron in them.

Below we show that the observed spin splitting anisotropy can be attributed to the effect
of SO coupling in the lateral electron motion in a narrow quantum well, and that its
values observed in various dot states in both samples A and B can be traced down to the
same SO coupling characteristics, namely, the effective SO-induced magnetic field
$B_{\mathrm{so}}=(%
\hbar m^{2})/(2e)(\varrho _{\mathrm{BR}}^{2}-\varrho _{\mathrm{D}}^{2})$.
For a quantum well lying in the (001) crystallographic plane of GaAs it is
convenient to choose coordinates along crystallographic directions $\mathbf{%
\hat{e}}_{1}=[110]$ and $\mathbf{\hat{e}}_{2}=[\bar{1}10]$, and to study the
effective 2D Hamiltonian in the form
\begin{eqnarray}
\mathcal{H}\! &=&\!\frac{\left( -i\hbar \nabla \mathbf{-A}-\mathbf{a}\right)
^{2}}{2m}+\dfrac{g}{|g|}\dfrac{\epsilon _{\mathrm{Z}}}{2}\mathbf{l\sigma }+V(%
\mathbf{r}),\;  \label{FullHam} \\
\mathbf{a} &\mathbf{=}&\frac{\hbar \sigma _{2}\mathbf{\hat{e}}_{1}}{2\lambda
_{1}}-\frac{\hbar \sigma _{1}\mathbf{\hat{e}}_{2}}{2\lambda _{2}}
\label{VectPot}
\end{eqnarray}%
where $\mathbf{A}=\tfrac{e}{2}B_{z}[\mathbf{r\times \hat{e}}_{z}]$, $%
\mathbf{l=B}/B$, $\mathbf{\sigma }=(\sigma _{1},\sigma _{2},\sigma _{3})$ is
the vector of Pauli matrices,
and $%
\epsilon _{\mathrm{Z}}=|g\mu _{\mathrm{B}}B|$ is the Zeeman energy
(in a 10nm wide GaAs/AlGaAs quantum well, $g<0$ according to
Snelling \textit{et al} \cite{Snelling}, and in the above
expressions we have already taken into account the negative sign
of the electron charge, so that $e>0$). The parameters $\varrho
_{\mathrm{D}}$ and $\varrho _{\mathrm{BR}}$ of the SO coupling
defined in Eq.~(\ref{Hso}) for a conventional choice of axes, $\mathbf{\hat{x%
}}=[100]$ and $\mathbf{\hat{y}}=[010]$ appear in the uniform non-Abelian vector potential
$\mathbf{a}$ in Eq.~(\ref{FullHam}) via the inverse of the
SO coupling length $\lambda _{1(2)}$ as $\lambda _{1}^{-1}=\left( \varrho _{%
\mathrm{D}}-\varrho _{\mathrm{BR}}\right) m$ and $\lambda _{2}^{-1}=-\left( \varrho
_{\mathrm{BR}}+\varrho _{\mathrm{D}}\right) m$, where $2\pi \lambda _{1(2)}$ characterizes
the distance at which spin precession of a polarised electron moving along
crystallographic direction $\mathbf{\hat{e}}_{1(2)}$ undergoes one complete revolution.

To analyze the case of a weak SO coupling for electrons bound in a
small-size quantum dot, $\lambda _{1,2}\gg \lambda $, we follow
Refs. \cite{AF,BF} and  perform a non-uniform unitary
transformation to the
electron wave function, $\psi (\mathbf{r})=U(\mathbf{r})\tilde{\psi}(\mathbf{%
r})$ and the Hamiltonian in Eq.~(\ref{FullHam}),
\begin{equation}
U=\exp \left\{ \tfrac{i}{2}\left[ \lambda _{1}^{-1}x_{1}\sigma _{2}-\lambda
_{2}^{-1}x_{2}\sigma _{1}\right] \right\} ,\;\mathcal{\tilde{H}}=U^{\dagger }%
\mathcal{H}U  \notag .
\end{equation}%
This transformation rotates locally the spin space by the angle
$R=\sqrt{\left( x_{1}/\lambda _{1}\right) ^{2}+\left(
x_{2}/\lambda _{2}\right) ^{2}}$ around the unit vector
$\mathbf{\hat{n}}\mathbf{=}\left( \lambda
_{1}^{-1}x_{1}\mathbf{\hat{e}}_{2}-\lambda _{2}^{-1}x_{2}\mathbf{\hat{e}}%
_{1}\right) /R$. As a result the coordinate frame for the electron
spin set in the center of the dot, $x_{1,2}=0$ gets adjusted to
the local orientation determined by the
SO-induced spin precession upon its displacement along the radius vector $%
\mathbf{r}$. The enrgy spectrum can be found from the transformed
Hamiltonian
\begin{equation*}
\mathcal{\tilde{H}}=\!\frac{1}{2m}\left( -i\hbar \nabla \mathbf{-A}-\mathbf{%
\tilde{a}}\right) ^{2}+\tfrac{1}{2}\epsilon _{\mathrm{Z}}\mathbf{\tilde{l}%
\sigma }+V(\mathbf{r}),
\end{equation*}%
where $\mathbf{\tilde{a}}=U^{\dagger }\mathbf{a}U+i\hbar U^{\dagger }\nabla
U $ and $\mathbf{\tilde{l}}(\mathbf{r})=(\mathbf{l\cdot \hat{n})\hat{n}+\hat{%
n}\mathbf{\times }\left[ \mathbf{l\times \hat{n}}\right] }\cos R-\mathbf{%
\left[ \mathbf{l\times \hat{n}}\right] }\sin R$. The transformation $U(%
\mathbf{r})$ aims to gauge out\ \cite{Gefen,MathurStone,AF,BF}
spin-orbit coupling, which appears in Eq.~(\ref{FullHam}) in the
form of uniform spin-dependent vector potential in
Eq.~(\ref{VectPot}). The latter goal cannot be achieved in full,
since Pauli matrices do not commute with each other. However, for
a weak SO coupling [$\lambda _{1,2}\gg \lambda $] and small
rotation angles $R\ll 1$, the residual $\mathbf{\tilde{a}}$ ,
\cite{BF}
\begin{equation}
\mathbf{\tilde{a}}=-\frac{[\mathbf{r}\times \mathbf{\hat{e}}_{3}]}{4\lambda
_{1}\lambda _{2}}\hbar \sigma _{3}+\hbar \lambda _{1,2}^{-1}\mathcal{O}%
[R^{2}],  \label{Aso}
\end{equation}%
is dominated \cite{Footnote1,footnote3} by the 'vector potential' of an effective magnetic
field which has the opposite sign for spin \textquotedblright up\textquotedblright\ and
\textquotedblright
down\textquotedblright\ electrons in the quantum dot, \cite{AF,BF}%
\begin{equation}
B_{\mathrm{eff}}=B_{z}-\sigma _{3}B_{\mathrm{so}}.  \label{BSO}
\end{equation}
It is this difference in the effective magnetic field seen by an
electron in the adjusted spin frame that causes the spin splitting
anisotropy. For the in-plane magnetic field orientation
($B_{z}=0$), the effective field $B_{\mathrm{eff}}=\pm
B_{\mathrm{so}}$ produces the same negligibly small 'diamagnetic'
shift $\sim B_{\mathrm{so}}^{2}$ in the orbital motion energy of
both spin components. Accordingly it does not alter the value of
the quantum well Zeeman splitting, $\Delta _{\Vert }=\epsilon
_{\mathrm{Z}}$.

For the perpendicular magnetic field orientation ($\mathbf{B}=B\mathbf{\hat{e%
}}_{z}$), the difference in $B_{\mathrm{eff}}=B_{z}\pm B_{\mathrm{so}}$ generates the
difference in the effective diamagnetic shift for two spin states and, therefore, an
additional energy splitting. It results in the anisotropy of spin splitting in the lowest
quantum dot state
\begin{equation}
\Delta _{\bot }-\Delta _{\Vert }= \dfrac{-g}{|g|} \dfrac{dE_{00}%
}{dB_{z}}2B_{\mathrm{so}}= \dfrac{-g}{|g|} \dfrac{\dfrac{\omega
_{c}}{2}\dfrac{e\hbar }{m}B_{\mathrm{so}}}{\sqrt{\omega ^{2}+\left( \tfrac{1%
}{2}\omega _{c}\right) ^{2}}}.  \label{AnisotropyFull}
\end{equation}%
Eq.~(\ref%
{AnisotropyFull}) is valid in both low and high magnetic field regimes. Its low-field
asymptotic
\begin{equation}
\Delta _{\bot }-\Delta _{\Vert }\approx \left( \dfrac{-g}{|g|}\right) B\hbar
e^{2}B_{\mathrm{so}}/(2\omega m^{2})  \label{AnisotropyLowField}
\end{equation}%
for $\omega _{c}<\omega $ also describes the situation of a strongly bound electron, such
as the resonance level A1. In the latter case, it is linear in the external field and
looks like the anisotropy of the Lande
factor. The high-field asymptotic of the result in Eq.~(\ref{AnisotropyFull}%
),
\begin{equation}
\Delta _{\bot }-\Delta _{\Vert }\approx \left( \dfrac{-g}{|g|}\right) \dfrac{%
e\hbar }{m}B_{\mathrm{so}}\;\;\mathrm{for}\;\omega _{c}\gg \omega ,
\label{AnisotropyHighField}
\end{equation}%
simultaneously describes the anisotropy of the spin splitting of the few lowest quantum
dot states $E_{0n_{-}}$. For all of them, the anisotropy energy
transforms into an offset with the sign dependent on the sign of $B_{\mathrm{%
so}}\propto (\varrho _{\mathrm{BR}}^{2}-\varrho _{\mathrm{D}}^{2})$ in Eq.~(%
\ref{BSO}) and on the sign of the electron g-factor.\\
Finally, we apply Eqs. (\ref{AnisotropyFull}-\ref{AnisotropyHighField}) to analyze the
peak splitting data shown in Fig.~\ref{fig:col}. Since in a 10nm-wide
GaAs quantum well the bare value of g-factor is negative \cite%
{Snelling}, larger values of spin splitting in a perpendicular field mean
that $B_{\mathrm{so}}\propto (\varrho _{\mathrm{BR}}^{2}-\varrho _{\mathrm{D}%
}^{2})>0$, pointing at the dominance of the Bychkov-Rashba term in
the SO coupling, presumably, due to the electron penetrating into
the AlGaAs barrier which is enhanced in a narrow quantum well.

The fit to the experimentally observed anisotropy produces the
values of the parameter $(e\hbar/m)\cdot B_{\mathrm{so}}$ which
are shown in  Table~\ref{tab1} and can be used to determine the
effective field $B_{\mathrm{so}}$ in each
sample. \ For sample B, the two extracted values of $B_{\mathrm{so}%
}=25\pm 6$~mT for B1 and $25\pm 8$~mT for B2-B4 were obtained using
Eq.~(\ref{AnisotropyHighField}) and agree to each other. The strongly confined state
observed in sample A [peak A1, Fig.~\ref{fig:col}(a)] was analyzed using the result in
Eq.~(\ref{AnisotropyLowField}), and as a result we extracted $B_{\mathrm{so}}=19\pm 1$~mT.
\begin{table}
\begin{center}
\begin{tabular}[t]{|c|c|c|} \hline \textbf{sample} & A & B \\ \hline
 peak & A1 &\hspace{0.9em} B1\hspace{2.em}\vline\hspace{1em} B2-B4\\ \hline
$\dfrac{e\hbar }{m}B_{\mathrm{so}}$ &  $32\pm 2$~$\mathrm{\mu eV}$ &  $43\pm
10$~$\mathrm{\mu
eV}$ \mbox{ }\vline \mbox{ }$44\pm 16$~$\mathrm{\mu eV}$ \\ \hline $B_{\mathrm{so}}$ & $19$~mT & $25$~mT  \\
\hline
\end{tabular}\label{tab1}
\end{center}
\caption{Experimental parameters of the spin-orbit characteristics for sample A and B, as
discussed in the text.}
\end{table}
The sign of SO coupling characteristics in $B_{\mathrm{so}}\propto (\varrho
_{\mathrm{BR}}^{2}-\varrho _{\mathrm{D}}^{2})$ suggests that in a narrow quantum well
 Bychkov-Rashba coupling is dominant - in contrast to wide quantum wells
investigated in Raman scattering\cite{Jusserand}. The extracted SO coupling is also
stronger than that estimated from the bulk Dresselhaus term,\ $2\gamma \hbar
^{-4}\sum_{ijk}\epsilon ^{ijk}p_{i}^{2}p_{j}s_{j}$, (where $\epsilon ^{ijk}$
is anti-symmetric tensor) using $\gamma =\left( 26\pm 5\right)$~$\mathrm{eV%
\mathring{A}}^{3}$ collected from references
\cite{Jusserand,WeakLoc}. For the quantum well width $w=10$nm, the
Dresselhaus mechanism alone would generate $B_{\mathrm{so}}$ which
reduces the Zeeman splitting and is of much smaller absolute
value, $\hbar ^{2}m/2\varrho _{\mathrm{D}}^{2}\approx 2\pi
^{4}m(\gamma /w^{2})^{2}\approx 12$~$\mathrm{\mu eV}$. Using this
estimate, we can deduce the strength of the Bychkov-Rashba
coupling in quantum wells in the samples B(A) as, at least,
$(\hbar ^{2}m/2)\cdot\varrho _{\mathrm{BR}}^{2}\sim
54(44)$~$\mathrm{\mu eV}$.

We acknowledge sample growth by A. F\"orster, H. L\"uth, V.
Avrutin and A. Waag. This work was partially supported by BMBF
(RH), DFG (RH),
 by EPSRC (VF), DARPA under the QuIST Program (BA), and by ARDA/ARO Quantum Computing Program (BA).


\begin{thebibliography}{99}
\bibitem{footnote3} The transformation $U(\mathbf{r})$ also turns the
direction of Zeeman field seen by an electron in the local spin-frame, $%
\frac{1}{2}\epsilon _{\mathrm{Z}}U^{\dagger }\left( \mathbf{l}\cdot \mathbf{%
\sigma }\right) U=\frac{1}{2}\epsilon _{\mathrm{Z}}\mathbf{\tilde{l}}\cdot
\mathbf{\sigma }$. Although a non-uniform $\mathbf{%
\tilde{l}}(\mathbf{r})$  generates the spin relaxation in dots \cite{AF,BF}, it does not
change electron spectra in the linear order in $\lambda /\lambda _{1,2}$, since
\begin{eqnarray*}
\langle n_{+}n_{-}|\mathbf{\tilde{l}}|n_{+}n_{-}\rangle &\approx &\mathbf{l}-%
\tfrac{1}{2}\left( \langle x_{1}^{2}\rangle \lambda _{1}^{-2}+\langle
x_{2}^{2}\rangle \lambda _{2}^{-2}\right) l_{z}\mathbf{\hat{e}}_{3} \\
&&\;-\tfrac{1}{2}\langle x_{2}^{2}\rangle \lambda _{2}^{-2}l_{2}\mathbf{\hat{%
e}}_{2}\mathbf{-}\tfrac{1}{2}\langle x_{1}^{2}\rangle \lambda _{1}^{-2}l_{1}%
\mathbf{\hat{e}}_{1}.
\end{eqnarray*}%
It is substantially represented by the first term since in GaAs quantum wells $\epsilon
_{\mathrm{Z}}/\omega _{c}=|g|m/m_{e}<\allowbreak
0.03 $ (here, $m_{e}$ is the electron mass in vacuum) and corrections to $%
\mathbf{l}$ produce much less effect than the anisotropy energy given by
Eqs.~(\ref{AnisotropyFull}-\ref{AnisotropyHighField}).
\end{thebibliography}
\end{document}